\documentclass[oribibl,runningheads]{llncs}

\usepackage{graphicx,lscape}
\usepackage{aas_macros} %AASTeX 52 macros for ADS BibTeX journals

\begin{document}

\title{Robust Machine Learning Applied to Terascale Astronomical Datasets}

\author{Nicholas M. Ball\inst{1} \and Robert J. Brunner\inst{1,2} \and Adam D. Myers\inst{1}}

\institute{Department of Astronomy, MC-221, University of Illinois, 1002 W. Green St., Urbana, IL 61801, USA\\ \email{nball@astro.uiuc.edu},\\ WWW home page: \texttt{http://nball.astro.uiuc.edu} \and National Center for Supercomputing Applications, MC-257, 1205 W. Clark St., Urbana, IL 61801, USA}

\maketitle

\begin{abstract} %%%70-150 words; is 155 minus 5 in footnote
We present recent results from the LCDM\footnote{Laboratory for Cosmological Data Mining; http://lcdm.astro.uiuc.edu} collaboration between UIUC Astronomy and NCSA to deploy supercomputing cluster resources and machine learning algorithms for the mining of terascale astronomical datasets. This is a novel application in the field of astronomy, because we are using such resources for {\it data mining}, and not just performing simulations. Via a modified implementation of the NCSA cyberenvironment Data-to-Knowledge, we are able to provide improved classifications for over 100 million stars and galaxies in the Sloan Digital Sky Survey, improved distance measures, and a full exploitation of the simple but powerful $k$-nearest neighbor algorithm. A driving principle of this work is that our methods should be extensible from current terascale datasets to upcoming petascale datasets and beyond. We discuss issues encountered to-date, and further issues for the transition to petascale. In particular, disk I/O will become a major limiting factor unless the necessary infrastructure is implemented.
\end{abstract}

\section{Introduction}

In the last decade, astronomy has changed from a data-poor to a data-intensive science, and it is important that the analyses of this flood of new data are able to maximally extract important information. Two fundamental analysis modes are (1) the fitting of parametrized physical models to data, perhaps in a Bayesian framework using existing knowledge as priors, and, of equal importance, (2) the {\it empirical} investigation of data to characterize it on its own terms. This second method enables one to approach the data with a minimum of pre-existing theoretical bias, and to perhaps make new discoveries that would not be made by simply fitting existing models. With current terascale and upcoming petascale datasets, the exploratory mode requires advanced methods of data mining to enable the full extraction of information from the data.

In extracting this information, the classification of, and measuring the distances to, objects in astronomical survey data provides a fundamental characterization of the dataset. It forms a vital first step in understanding both its ensemble properties and those of individual objects, and leads to improved investigations of the astrophysical nature of the Universe.

In this paper, we present results from the Laboratory for Cosmological Data Mining, a partnership between the Department of Astronomy and the National Center for Supercomputing Applications (NCSA) at the University of Illinois at Urbana-Champaign (UIUC), in collaboration with the Automated Learning Group (ALG) at NCSA, and the Illinois Genetic Algorithms Laboratory at UIUC. Such a collaboration between institutions is atypical, and the combination of expertise allows us to apply the full power of machine learning to contemporary terascale astronomical datasets, guided by appropriate science drivers on which we are then able to follow up.

In the following, we first describe the data, to place the task in context, followed by the computing environment, a brief overview of the results thus enabled, the issues raised by the process, and prospects for the extension of this work to the petascale.

\section{Data}

The astrophysical objects considered here are stars, galaxies and quasars. The stars are a subset of the approximately $10^{11}$ to $10^{12}$ stars contained within our Galaxy. Likewise the galaxies are a subset of the approximately $10^{11}$ to $10^{12}$ galaxies in the observable Universe. Quasars are a phenomenon at the heart of many galaxies in which matter is being accreted onto a supermassive black hole. Whereas galaxies can be thought of as agglomerations of stars, quasars are driven by different physics. Their brightness renders them visible throughout most of the observable Universe, to much greater distances at the same photometric limit than most galaxies. They are thus an important tracer of cosmic structure and evolution.

Modern sky surveys typically contain a set of objects with spectra and a much larger superset with just photometry. The spectra give detailed information on the type of object seen and, for a galaxy or quasar, its distance via the redshift. The photometry, however, just gives the brightness of the object in a few broad wavebands. Therefore, the objects with spectra, provided they form a reasonably representative subsample, can be input as a training set to a supervised machine learning algorithm. The spectral information forms the target property (e.g., type or redshift), and the photometric information forms the training features. The machine learning can then be used to make empirical predictions about the much larger sample set objects for which only photometry is available, without a priori sample bias introduced by, for example, color, formed by the difference in magnitudes.

The data investigated here are obtained from the Sloan Digital Sky Survey (SDSS) \cite{york:sdss}. This is a project to digitally map $\pi$ steradians of the Northern Galactic Cap in five broad optical photometric bands, $u$, $g$, $r$, $i$, and $z$. The survey is unprecedented in its combination of accurate digital photometry, numbers of objects and spectra, and is thus the best dataset currently available for studying the ensemble properties, classifications, and redshifts of astrophysical objects.

We consider the Third and Fifth Data Releases of the survey (DR3 and DR5). DR3 consists of 142,705,734 unique objects, of which 528,640 have spectra. There are 6.0 TB of images, and object catalogs in either FITS \cite{wells:fits} format (1.2 TB) or as a SQL database (2.3 TB). DR5 is a superset of DR3, approximately one and a half times the size.

We utilize the objects with spectra as training sets. Here, the training features are the four colors $u-g$ through $i-z$ in the four magnitude types measured by the SDSS, although it is trivial to add any other feature that is measured for objects within the SDSS, for example, measures of morphology. For each dataset we perform {\it blind tests} on subsets of the objects with spectra to obtain a realistic measure of the performance of the machine learning algorithm.

\section{Computing Environment}

The machine learning algorithms are implemented within the framework of the Data-to-Knowledge toolkit (D2K) \cite{welge:d2k}, developed and maintained by the ALG at NCSA. This allows the straightforward implementation of numerous $\mathrm{Java^{TM}}$ modules, connected together via a GUI interface to form itineraries. The machine learning algorithms available include decision tree, $k$-nearest neighbor ($k$NN), genetic algorithm, artificial neural network, support vector machine, unsupervised clustering, and rule association. For the terascale datasets in use here, our implementation includes enhanced versions of the standard modules that stream data of a fixed type, for example single-precision floating point. The standard D2K datatype, table, is more flexible, and there are many tools to deal with tables, but the operations usually require the full dataset to fit into machine memory, which is not an option for our analysis.

The algorithms are run on the Xeon Linux Cluster \textit{Tungsten} at NCSA, as part of a peer-reviewed, nationally allocated, LRAC allocation to the LCDM project on this and other machines, renewed over multiple years. Tungsten is a distributed memory system composed of 1280 compute nodes, 2560 processors, 3.8 TB ECC DDR SDRAM total memory, and a peak performance of 16.4 TFlops. Each node is a Dell PowerEdge 1750 server running Red Hat Linux with two Intel Xeon 3.2 GHz processors, 3 GB of memory, a peak double-precision performance of 6.4 Gflops, and 70 GB of scratch disk space. A further 122 TB of general scratch disk space is available via the Lustre filesystem, and the archival file storage is provided by the 5 PB UniTree DiskXtender mass storage system, connected via FTP interface. The node interconnect is Myrinet 2000. Batch jobs are invoked via the Load Share Facility (LSF).

In the publicly distributed version of D2K, parallelism is implemented via the installation of the D2K Server on the intended compute nodes, whose locations are specified in advance. On Tungsten, this approach is unsuitable because different nodes are used each time the overall batch job is executed. What D2K does allow, however, is invocation in batch mode via a script. We therefore invoke the itineraries in this way directly via an LSF shell script on a head node. The script uses ssh to connect to multiple child nodes. Each node runs an independent instance of the itinerary on a subset of the data and/or the machine learning parameter space. The input data are accessed from the global scratch disk space and is not altered, and the output data are independent in the sense that they only need to be collated when the all of the child nodes have returned their results. Our approach is therefore an example of task farming, with no required communication between child nodes. Thus communication libraries such as MPI are not needed.

\section{Science Results}

We give a brief overview of the astrophysical results enabled by the computing environment described above.

Using decision trees, we \cite{ball:dtclassification} assigned probabilities P(galaxy, star, neither-star-nor-galaxy) to each of the 143 million objects in the SDSS DR3. This enables one to, for example, either emphasize completeness (the fraction of the true number of targets correctly identified), or efficiency (the fraction of targets assigned a given type that are correct) in subsamples, both of which have important scientific uses. Our study was the first such classification performed on an astrophysical dataset of this size.

The upper panel of Fig. 1 shows a result typical of those in the literature, prior to that from LCDM, for photometric versus spectroscopic redshift for quasars, here reproduced by blind-testing 20\% of 11,149 SDSS DR5 quasars using a single nearest-neighbor (NN) model. Most objects lie close to the ideal diagonal line, but there are regions of `catastrophic' failure, in which the photometric redshift assigned is completely incorrect (more than 0.3 from the true redshift). The correct assignation of these redshifts is vital for large numbers of future studies of the large scale structure in the universe, and the use of NN enables us to significantly reduce the instance of these catastrophics, by `pulling in' the outliers (Fig. 1 of \cite{ball:ibphotoz}). The RMS dispersion between the photometric and spectroscopic redshift is reduced by 25\%, from $\sigma \sim 0.46$ to $\sigma = 0.343 \pm 0.005$ (error via 10-fold cross-validation).

\begin{figure}
\centering
\includegraphics[width=4.3in]{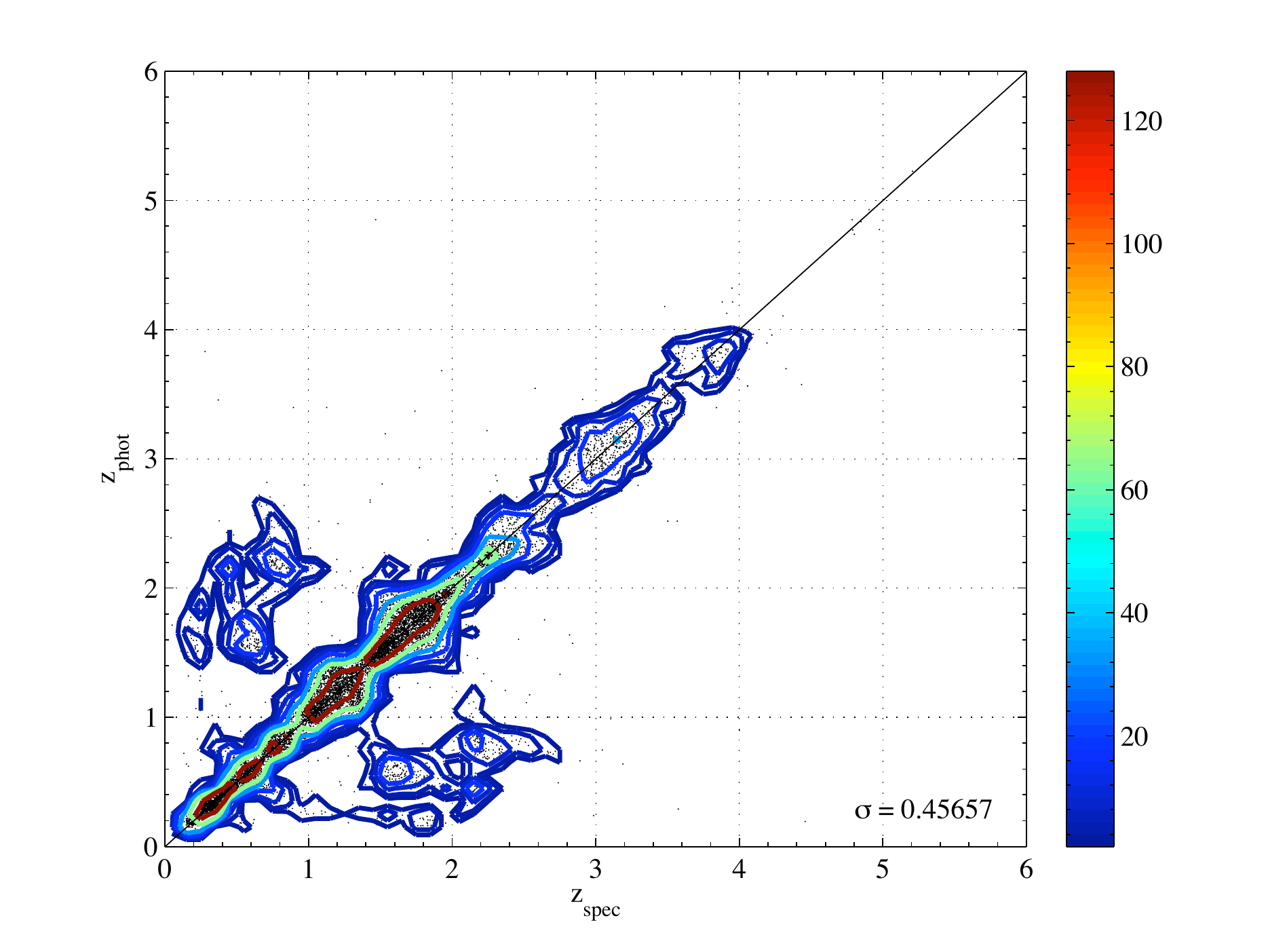}
\includegraphics[width=4.3in]{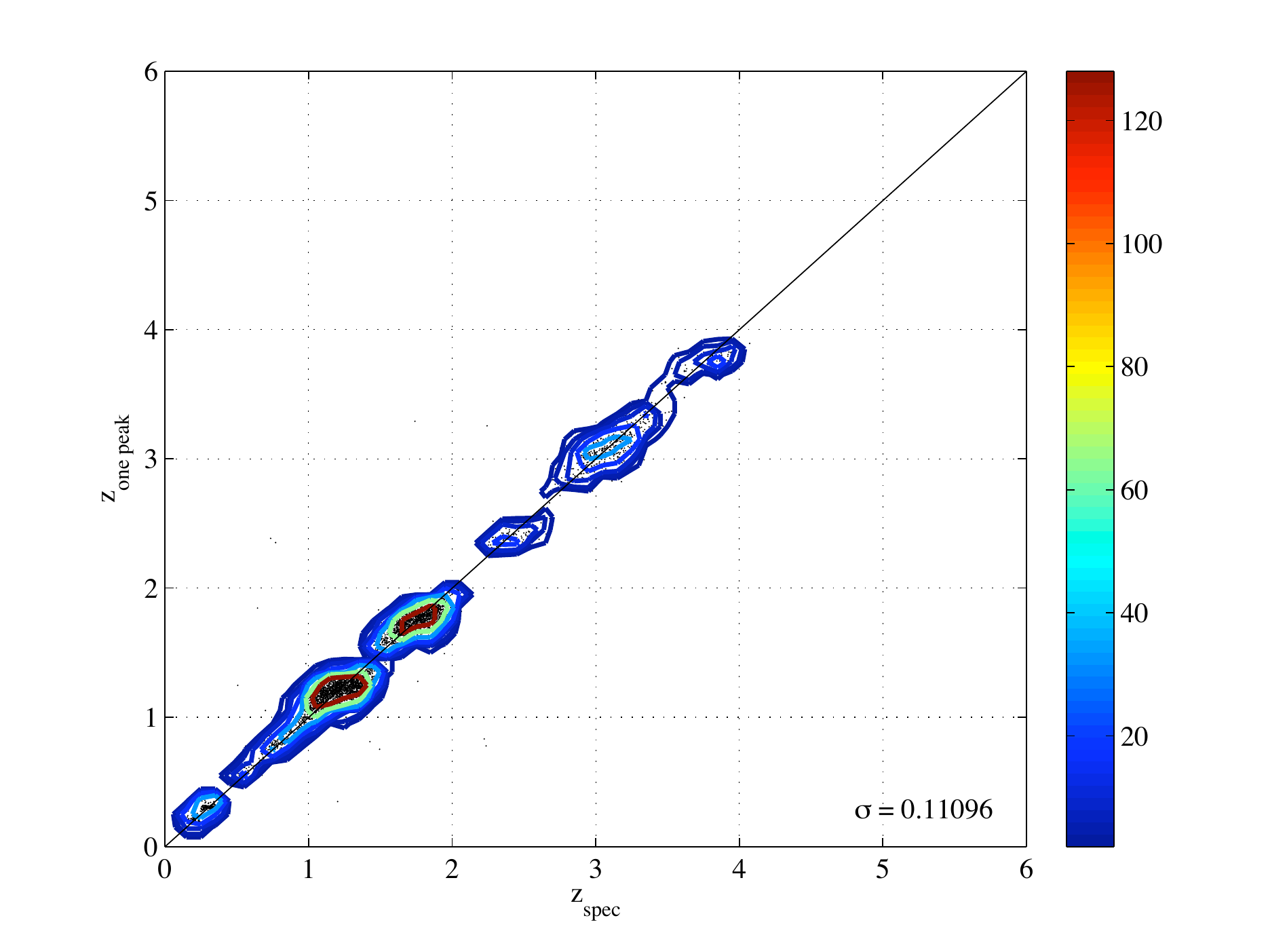}
\caption{Spectroscopic versus photometric redshift for Sloan Digital Sky Survey Data Release 5 quasars. The upper panel shows a result typical of those in the literature prior to our group's work. The lower panel shows the result of using machine learning to assign probability density functions, then taking the subset of objects with a single peak in probability. This virtually eliminates the regions of `catastrophic failure', a result of vital importance for many areas of cosmology.}
\end{figure}

However, as we show in \cite{ball:pdfphotoz}, we can improve on this even further: because every observed object magnitude in the SDSS has an associated error, one can perturb the testing set numerous times according to the errors on the input features, and use the resulting variation to generate full probability density functions (PDFs) in redshift. This solves a classic machine learning problem of how to account for varying errors on the training set inputs. Taking the mean value from each PDF gives a similar RMS dispersion to the result of \cite{ball:ibphotoz}, $\sigma \sim 0.35$, as expected. However, for the subset of quasars which have a single PDF peak, about 40\% of the objects, the RMS dispersion between the photometric and spectroscopic redshift is vastly improved to $\sigma = 0.117 \pm 0.010$. This reduces the fraction of catastrophics from 20.2\% to 0.7\%, essentially eliminating them from the sample. This is shown in the lower panel of Fig. 1. The panel shows only the first of the ten cross-validations, and thus the value of $\sigma$ is within the error of, but not exactly equal to, 0.117.

\section{Computing Results}

Here we summarize Tungsten resource usage. Because this is a scientific application using existing algorithms on existing proven hardware, we did not keep detailed quantitative records of quantities such as overall efficiency, time spent in compute, communication, or waiting on data from local or remote filesystems. We monitored the performance with the aim of having the tasks completed in a reasonable amount of time compared to that available, making sure, of course, that we were not na\"ively wasting time. This is likely to be typical of many science applications in the future. Table 1 shows some typical run times for the creation of these results.

What we can say is that, generally, the usage of the resources is likely to be efficient, because it is dominated by the processing of data in situ at each node. Communication between nodes is negligible because we are task farming, and at present we are not I/O limited. One of the main inefficiencies is that the time taken for the job is that taken by the slowest node, because the output requires the full set of results to be collated. The full set of nodes used for the job is reserved during the run. The problem is most acute when working with subsets of machine learning parameters, because different parameters can result in significant differences in runtime that can be difficult to predict.

For decision tree, in training, the data are small but the work is large, so this is CPU dominated. In applying the tree to 143 million objects, the 80G file describing the objects is streamed through the D2K itinerary, but the classification process is fast (about 2000 objects per second), so it is possible that data access here is more significant. For NN and $k$NN, the training is trivial: memorize the positions of the objects in parameter space, and the training data are again small. The significant computation therefore occurs for the application to new objects, and scales as $sp$ for $s$ training set objects and $p$ testing set objects, i.e. it is $O(n^2)$. Thus the run time is dominated by the processing time to apply the algorithm to the new data, and the available parallelism.

\section{Terascale Issues}

We now describe some of the issues encountered during data mining on Tungsten. These form a case study of the specific issues encountered in the course of our work, but they are of general relevance in working with terascale data, and show the in-practice implementation of a science-driven data mining application on a high performance computing cluster.

\begin{itemize}
\item We are using tens to hundreds of parallel nodes and streaming many GB of data. D2K must therefore be invoked via batch script, negating the advantages of its GUI interface. The lack of an integrated cyberenvironment results in batch scripts that contain many tens of settings, manually fixed file locations, and commands, that are prone to error. Our large scale embarrassingly parallel processing of data via task farming is likely to become an increasingly common scientific application. A simple task farming utility that can be called via the job submission script would help to simplify these scripts and make them more robust.
\item Batch job submission via LSF is inflexible, subject to fixed wallclock times and numbers of nodes, unpredictable queuing times and no recourse if a job fails, perhaps due a bug in the script, or a hardware problem in any of the nodes or the file system. The ability of the system to perform application checkpointing and to check for failure in any of the tasks on the child nodes, and rerun that task if such failure is encountered, would be another useful advance.
\item Any dataset larger than that which fits comfortably in the Tungsten home directory (5G quota) must be stored on the Unitree mass storage system, because the scratch filesystem is purged. Unitree is occasionally subject to outages in access or significant wait times. In combination with the queuing system for batch jobs described above, this can make new scripts that are to access large datasets time-consuming to debug. This is because the data must be either accessed directly from Unitree, extracted to the scratch disk before debugging, or subsampled to a manageable size.
\item In general, machine learning algorithms have a large parameter space of adjustable settings, which inevitably require significant computing time to optimize before useful results are generated. The nearest-neighbor approach mitigates this to some extent, in particular the single nearest neighbor, but this in turn is computationally expensive due to the large number of distance calculations required for each object.
\item The present lack of fainter training data forces us to extrapolate in order to classify the whole SDSS. While the data and results we obtain are well-behaved (our training features, the object colors, are largely consistent beyond the limit of the spectroscopic training set), it will always be the case in astronomy that some form of extrapolation is ultimately required. This is simply due to the fact that, physically, photometry to a given depth requires a shorter observation time than spectroscopy. Thus, while our supervised learning represents a vital proof-of-concept over a whole terascale survey, and our already good results can only improve for a given survey as more training data become available, ideally the general data mining approach should include semi-supervised or unsupervised algorithms. These will help to fully explore the regions of parameter space that lie beyond the available training spectra.
\item The data size is such that integrating the SQL database with D2K via JDBC is impractical, and the data must be stored as flat files. As database engines become more sophisticated, however, it could in the future become possible to offload partial or entire classification rules to a database engine. Doing so, however, would require supercomputing resources for the database engine, which would result in an entirely new class of challenges.
\item Many algorithms exist for the implementation of data mining on parallel computer systems beyond simple task farming (e.g., \cite{freitas:parallel,kargupta:dpkd,zaki:parallel}), but these are not widely used within science compared to the commercial sector, and most scientists are not trained to fully exploit them on parallel computing systems. Thus interdisciplinary collaboration and training between scientists and experts in data mining and supercomputing will continue to be essential, and will only increase in importance on the petascale.
\item In general, most existing tools designed to deal with data mining require that the data fits in the machine memory. What is needed is a cyberenvironment in which the streaming of data that is larger than machine memory is fully integrated into the system, to the extent that the file size does not affect the top-level workflow seen by the user.
\end{itemize}

\section{Towards Petascale Computing}

Given the petascale datasets planned for the next decade (e.g., the Large Synoptic Survey Telescope, scheduled for first light in 2014, will generate around 7 PB per year, compared to the 10 TB of the SDSS), it is vitally important that contemporary data mining can be carried out successfully on this scale. We discuss some of the issues that we expect to encounter, and estimate our scalability from terascale to petascale.

\begin{itemize}
\item For many applications on the petascale, the performance becomes I/O limited. This is quantified by \cite{bell:petascale}, who apply Amdahl's law \cite{amdahl:law} that one byte of memory and one bit per second of I/O are required for each instruction per second, to predict that a petaflop-scale system will require one million disks at a bandwidth of $100~{\mathrm{MB~s^{-1}}}$ per disk. They also state that data should be stored locally (i.e., not transferred over the Internet), if the task requires less than 100,000 CPU cycles per byte of data. Many contemporary scientific applications are already such that local storage is favored by over an order of magnitude.
\item We can estimate the sizes of the upcoming datasets compared to the required computing power: in 3--5 years, the SDSS will have completed its primary data releases, which will total less than twice the data available now. However, the proposed Petascale System {\it Blue Waters} will come online at NCSA in $\sim 5$ years, increasing the computing power by $O(100)$ compared to Tungsten. In ten years, the data size will be $O(1000)$ times that now, i.e., $O(10{~\mathrm{PB}})$ versus $O(10{~\mathrm{TB}})$. However, this will be a further $\sim 5$ years on from Blue Waters, which, if Moore's Law continues to hold, will be another factor of 10 in computing power. Thus computing power will have increased by $O(1000)$ times to match the data.
\item A similar improvement in affordable disk storage is predicted by \cite{szalay:petabyte}. We therefore predict that it will be possible to perform data mining on petascale astronomical datasets, {\it provided} that (a) the I/O infrastructure exists to match the processing power and (b) the task farming model scales to many thousands of processors. Condition (a) can be met so long as the necessary investment is made, and (b) is met because the task is embarrassingly parallel. We emphasize that this I/O infrastructure is vital to enable the continued effective use of scientific data. The million disks of \cite{bell:petascale} are required even if the I/O rate scales as well as they predict, and this is by no means certain in a real-world system.
\item In our case, if we were to continue to use NN, the situation is in fact a little more complex, because the algorithm scales as $sp$, ($s$ spectroscopic objects in the training set, $p$ photometric objects to classify). Currently we have $s = O(10^6)$ and $p = O(10^8)$, and in the next decade these are expected to increase to $s = O(10^7)$, via spectrographs such as WFMOS or KAOS, and $p = O(10^{10})$ via, e.g., the LSST. The less rapid growth of $s$ and the dominance of $p$ that only grows by $O(100)$\footnote{The final factor of 10 will come from from multiple observations in the time domain; while time domain astrophysics will become an important new research area, a lot of work will likely use a single stacked instance for each object.} gives an increase in $sp$ of $O(1000)$. So our mode of data analysis is viable for at least the next decade. Nevertheless, each object may be run on an unlimited number of NN models, and $sp$ is still an $n^2$ algorithm, like many of those in data mining. Therefore, there is incentive to pursue both even faster processing, particularly for distance calculations between pairs of points in parameter space, in the form of field-programmable gate arrays (FPGAs), general purpose graphical processing units (GPGPUs), Cell processors, etc. Similarly, faster I/O is always desirable. Alternatively, there are treelike NN implementations or other algorithms such as decision tree that scale as $n$~log~$n$, at the expense of losing information.
\item As mentioned, conventional hardware, in the form of large clusters of multicore compute nodes, is in principle scalable to the petascale. However, FPGAs, GPGPUs, and Cell processors may be more suited to many data mining tasks, due to their embarrassingly parallel nature and ability to rapidly perform a large number of distance calculations. The LCDM group, in collaboration with the Innovative Systems Laboratory at NCSA, has demonstrated a 100x speedup for the two-point correlation function (another astronomical statistic with a large number of distance calculations that scales na\"ively as $n^2$) on FPGAs using an SRC-6 MAP-E system \cite{brunner:fpganpcf}. A preliminary implementation of a $k$NN algorithm has also been made, although it was non-trivial because the algorithm itself had to be altered (parallelized), in addition to wrapping the code.
\end{itemize}

\section*{Acknowledgments}

The authors acknowledge support from NASA through grants NN6066H156 and NNG06GF89G, from Microsoft Research, and from the University of Illinois. The authors made extensive use of the storage and computing facilities at the National Center for Supercomputing Applications and thank the technical staff for their assistance in enabling this work. The authors also wish to thank the reviewers for their helpful suggestions to improve the presentation and technical content of this paper.

%\bibliographystyle{/Users/nball/latex/bst/splncs}
%\bibliography{/Users/nball/Documents/latex/refs/refs}

%Manually added space after \aj etc.

\begin{landscape}
\begin{table}
\caption{Summary of resource usage to obtain the results presented. DT = decision tree, and one DT model may include multiple bagging and cross-validation. NN = nearest neighbor algorithm, and $k$ is the number of nearest neighbors employed. The time taken is the number of seconds the job ran for multiplied by the number of nodes. The number of objects per second is the number of objects divided by this time. Starred values are estimated. Note that many of the individual runs are not especially large, but that the sum total implementation of the framework and the data storage would be infeasible on a desktop machine}
\begin{tabular}{llllllllc}
\hline\noalign{\smallskip}
Dataset &Objective &Task &Algorithm &No. Objects &No. models &Time taken / s &Objects / $s^{-1}$ &Reference\\
\noalign{\smallskip}
\hline
\noalign{\smallskip}
SDSS DR3 &Classification &Training     &DT            &381,654     &6868* &27,274,395* &96*   &\cite{ball:dtclassification}\\
SDSS DR3 &Classification &Blind test   &DT            &95,413      &1     &1601        &60    &\cite{ball:dtclassification}\\
SDSS DR3 &Classification &photoPrimary &DT            &142,705,734 &1     &71,718      &1990  &\cite{ball:dtclassification}\\
SDSS DR5 &Photoz         &Training     &$k$NN, $k=22$ &44,597      &5000  &40,000*     &5000* &\cite{ball:ibphotoz}\\
SDSS DR5 &Photoz         &Blind test   &$k$NN, $k=22$ &11,149      &5000  &1,227,730   &45    &\cite{ball:ibphotoz}\\
SDSS DR5 &Photoz PDF     &Training     &NN ($k=1$)    &44,597      &1     &10*         &5000* &\cite{ball:pdfphotoz}\\
SDSS DR5 &Photoz PDF     &Blind test   &NN ($k=1$)    &11,149      &100   &34,723      &32    &\cite{ball:pdfphotoz}\\
\hline
\end{tabular}
\end{table}
\end{landscape}

\end{document}